\def\be{\begin{equation}}
\def\ee{\end{equation}}
\def\bea{\begin{eqnarray}}
\def\eea{\end{eqnarray}}
\def\bq{\begin{quote}}
\def\eq{\end{quote}}

\def \lsim{\mathrel{\vcenter
     {\hbox{$<$}\nointerlineskip\hbox{$\sim$}}}}
\def \gsim{\mathrel{\vcenter
     {\hbox{$>$}\nointerlineskip\hbox{$\sim$}}}}

\def\gappeq{\mathrel{\rlap {\raise.5ex\hbox{$>$}}
{\lower.5ex\hbox{$\sim$}}}}

\def\lappeq{\mathrel{\rlap{\raise.5ex\hbox{$<$}}
{\lower.5ex\hbox{$\sim$}}}}

\def\bbz{fa Z \kern-8.9pt Z}

\def\ga{\mathrel{\raise.3ex\hbox{$>$\kern-.75em\lower1ex\hbox{$\sim$}}}}
\def\la{\mathrel{\raise.3ex\hbox{$<$\kern-.75em\lower1ex\hbox{$\sim$}}}}
\def\m12{m_{1\!/2}}
\def\tb{\tan\beta}
\def\ohsq{\Omega_{\widetilde\chi}\, h^2}
\def\mchi{m_{\tilde \chi}}
\def\gev{{\rm \, Ge\kern-0.125em V}}
\def\tev{{\rm \, Te\kern-0.125em V}}

\documentstyle [12pt,epsfig]{article}
\begin{document}
\thispagestyle{empty}
\begin{flushright}
{CERN-TH/99-170} \\
{MADPH-99-1124}\\
{hep-ph/9907365}\\

\end{flushright}
\vspace{1cm}
\begin{center}
{\large Dark Matter Abundance  and Electroweak 
 Baryogenesis }\\
{ in the CMSSM } \\
\vspace{.2cm}
\end{center}
\vspace{1cm}
\begin{center}
{Sacha Davidson }\\
{CERN Theory Division,
CH-1211 Gen\`eve 23, Switzerland}\\
\vspace{.5cm}
{Toby Falk }\\
{ Department of Physics, University of Wisconsin, Madison, WI~53706,
USA}\\
\vspace{.5cm}
{Marta Losada}\footnote{On leave of absence from the Universidad Antonio Nari\~{n}o, Santa Fe de Bogot\'a, COLOMBIA.}\\
{CERN Theory Division,
CH-1211 Gen\`eve 23, Switzerland}
\end{center}
\hspace{3in}

\begin{abstract}
The Minimal Supersymmetric Standard Model has a 
candidate dark matter particle in its spectrum, and may 
be able to generate the baryon asymmetry of the Universe (BAU) at the electroweak phase
transition. In the Constrained MSSM, we find the area
of parameter space which is allowed by accelerator
and precision tests, which produces a relic dark matter
abundance in the observationally favored
window $0.1 \lappeq \Omega h^2 \lappeq 0.3$, and where
baryon plus lepton number violating processes
are out of equilibrium after the electroweak
phase transition.
\end{abstract}

\vspace{1cm}
{June 1999} \\

\newpage

\section{Introduction}

Two intriguing features of the Minimal Supersymmetric Standard Model
(MSSM) from a cosmological perspective are that the Lightest
Supersymmetric Particle (LSP) is a stable dark matter \cite{DM} candidate,
and that electroweak baryogenesis\cite{EWB} might be possible for
certain parameter choices. Moreover, large regions of the baryogenesis
parameter space  should be accessible to the presently running
LEP200.

 The stability of the LSP is ensured by
$R$-parity, which is often imposed on supersymmetric models to avoid
rapid proton decay and to satisfy other phenomenological bounds 
\cite{Rp}.  The $R-$parity of a particle is $(-1)^{3B+L+2S}$, where
$S$ is its spin. Standard Model particles are even  under $R-$parity, and supersymmetric
partners are odd, so the LSP  cannot decay.  
The LSP freezes out of chemical equilibrium
in the early universe, enhancing its relic abundance, and its mass
density is often large enough to be cosmologically
significant, particularly when the LSP is a gaugino-type
neutralino.  It is an appealing feature of supersymmetry that it
naturally provides a stable dark matter candidate, with a relic
density in the cosmologically preferred range $0.1\lappeq \Omega h^2 
\lappeq 0.3$, discussed below.

 A second cosmologically interesting feature of
supersymmetric models is that it may be possible
to generate the baryon asymmetry at the Electroweak
Phase Transition (EWPT) \cite{EWB}. 
The three ingredients required for baryogenesis, Baryon number violation, C and CP violation
and out-of-equilibrium dynamics \cite{sak},
are present in the Standard Model (SM) at the EWPT.   However, generating
the observed baryon asymmetry at the EWPT is not
possible in the Standard Model \cite{fs, gavelaetal} because the
non-perturbative Baryon
plus Lepton ($B+L$) number violating sphaleron processes
\cite{sphal} remain in equilibrium after  the
phase transition  and will wash out any
asymmetry in $B+L$ present at that time \cite{KetalSM}. 
A necessary condition for electroweak baryogenesis is
that these processes
be out of equilibrium after the $B+L$ asymmetry
is generated at the phase transition. This requirement translates into having a sufficiently
strong first order electroweak phase transition at finite temperature and
 is satisfied for various
extensions of the SM \cite{marta1}, including the MSSM
with a light stop \cite{CQW,KetalMSSM,Losada4}.

There has been much work on constraining
supersymmetric models using the combined
bounds from accelerator experiments and the cosmological
relic abundance \cite{tf1}. The area
of SUSY parameter space in which
electroweak baryogenesis could work
(no $B+L$ violation after the transition)
has also been studied in some detail  \cite{SUSYBL4,KetalMSSM,CQW}.
The purpose of this paper is to combine
these two cosmologically interesting features
of supersymmetry. 
In the Constrained
MSSM (CMSSM), where the soft SUSY
breaking parameters $m_0, m_{1/2}$ and $A$ 
are universal at the GUT scale, we find
the area of parameter space  consistent
with present LEP bounds on sparticle masses, with
the measured $b \rightarrow s \gamma $ branching ratio,  and
with precision tests of the $\rho$ parameter,
where it would be possible to preserve a
baryon asymmetry generated at the
electroweak phase transition,
and where the relic density of LSPs 
satisfies
$0.1 \lappeq \Omega h^2 \lappeq $0.3.

It is interesting that there remains a finite area of
CMSSM parameter space satisfying all the  constraints, although the
area is not very  large (see Figs.~\ref{fig:ma12.45n}-\ref{fig:xtb}). 
Moreover, most 
of the area is accessible to LEP200; Higgs searches should tell us
if we live in this area of CMSSM parameter space.

\section{Cosmological and Experimental Constraints}

A prerequisite for successful electroweak baryogenesis 
is that any asymmetry created at the phase transition
not be washed out afterwards. This means that
the $B+L$ violating
electroweak processes, which are suppressed
by a factor of order $ e ^{- 4 \pi \phi/(g T)}$
after the transition,
 must be out of equilibrium. This  will
be the case if \cite{KRS}
\be
\frac{\phi(T_c)}{T_c} \gsim 1, \label{1}
\ee
where $\phi(T_c)$ is the vacuum expectation value (vev)
of the Higgs in the broken minimum of the Higgs
potential at the temperature $T_c$ where
the two minima are degenerate.

The jump in the Higgs vev can be calculated
in various ways. One can compute
loop corrections in perturbation theory to the
finite temperature effective
potential $V_{eff}(\phi,T)$. The generic form of the effective potential
potential is

\be
V_{eff}(\phi, T) = \frac{\gamma}{2} T^2 \phi^2 
\left[1 - B  \ln \left(\frac{\phi}{T} \right)\right]
- \frac{m^2}{2} \phi^2
 - E T \phi^3 + \frac{\lambda}{2} \phi^4 \label{V},
\ee
in which case, at the critical temperature,
\be
\frac{\phi(T_c)}{T_c} =  \frac{E}{2 \lambda} +  \frac{1}{2}
\sqrt{  \frac{E^2}{ \lambda^2} + \frac{2 \gamma B}{\lambda}} 
    \simeq \frac{E}{\lambda}\left( 1 + 
\frac{\lambda \gamma B}{E^2} \right) ~~. 
\label{?}
\ee
For a given model, the parameters $\gamma$, $E$, 
$\lambda$ and $B$ are calculable functions of the 
zero temperature  coupling constants.
The $B$-term is an approximation aimed to model
two-loop QCD corrections in the MSSM \cite{Esp}.
In this perturbative approach, 
the ratio of the Higgs vev to the temperature is increased by having a larger cubic 
term, a larger $B$, and a smaller quartic
coupling $\lambda$. 
At zero-temperature, $\lambda$ is
related to the mass of the Higgs particle.
At tree level in the SM, $m_H^2 = 4 \lambda \phi^2$,
so small $\lambda$ corresponds to small
Higgs mass. When quantum corrections
are included, the relationship becomes
less simple, but the condition (\ref{1})
still translates into an upper bound on
the Higgs mass.

The cubic term  arises  in the high
temperature expansion of the one-loop contribution 
to the effective
potential from
bosons which couple to the Higgs vev \cite{AE}. There is a term
$-(m_b^2)^{3/2}T/(12 \pi)$, which
contains a term  $\phi^3 T$  
for a boson mass $m_b^2 = m^2 + c_1 \phi^2 + c_2 T^2$.
If $m^2$ and $c_2$ are small enough,
the effective potential is approximately of the form (\ref{V}). 
The one-loop  contribution to $E$ from the $W$ and $Z$ bosons
 is $(2g^3 + [\sqrt{g^2 + g^{'2}}]^3)/(32 \pi)$ \cite{ESM}, 
but this is not large enough to satisfy equation
(\ref{1}) for experimentally allowed
 values of the Higgs mass (which determines $\lambda$)
in the SM.

In extensions of the Standard Model, additional particles in the
thermal bath contribute to the effective potential.  Of particular interest are
the new bosons which  couple strongly to the Higgs field
and can modify the contributions to the cubic term in the potential.
In the MSSM \footnote{ For large values of $m_{A}$, the lightest Higgs quartic self-coupling is fixed at tree-level to be
$\lambda = {g^2 \over 8} \cos^2 2\beta$.}, a light squark would contribute a
term $\sim   h_q^3 \sin^3 \beta/(4 \pi \sqrt{2}) $
to $E$, which would be 
substantial for the  top Yukawa coupling
$h_t$.  To see this, consider the finite temperature
mass matrix for the doublet  and singlet  
stops:
\be
\left[
\begin{array}{cc}
m^2_{Q_3} + \frac{1}{2}h_t^2 \sin^2 \beta ~\phi^2 +\frac{1}{8} g^2 \cos 2 \beta~
\phi^2 + {\cal O}( T^2) &  -\frac{1}{\sqrt{2}} h_t \phi \sin \beta (A + \mu \cot \beta ) \\
 -\frac{1}{\sqrt{2}} h_t \phi  \sin \beta (A + \mu \cot \beta )
 & m^2_{U_3} + \frac{1}{2}h_t^2 \sin^2 \beta ~\phi^2 +  {\cal O}( T^2)
\end{array}
\right] \label{AAA}
\ee
where we normalize $m_W = g \phi/2$, we neglect
the $U(1)$ coupling,  
and $m_{Q_3} (m_{U_3})$ is the third generation doublet
(singlet) squark soft SUSY-breaking mass.
If we neglect doublet-singlet mixing, then 
$(m^2_{\tilde{t}_R} )^{3/2} = (m^2_{\tilde{U}_3} + h_t^2 
\sin^2 \beta ~\phi^2/2
+4 g_s^2 T^2/9 + \frac{1}{6}(1 + \sin^2 \beta)h_t^2 T^2 )^{3/2}$ is of order
$h_t^3 \phi^3$, if $m^2_{\tilde{U}_3}$ is negative and cancels the finite temperature contributions.
We cannot make the other stop light,
because two light stops increase the $\rho$
parameter beyond its measured value.
In the MSSM at two loops,
the  light singlet stop contribution to 
$E$  is sufficient to satisfy equation (\ref{1}).

 A drawback
to the perturbative evaluation of the effective potential
is that it has infra-red divergences 
as the Higgs vev $\phi$ approaches zero. These are due
to bosonic modes whose only mass in the perturbative calculation
is proportional to  $\phi$. Resummation is then employed
to deal with these divergences. The net effect in the case
of the Standard Model is to reduce the contribution from
the gauge bosons to the cubic term, thus weakening the strength of
the phase transition.  An alternative way to compute the ratio
(\ref{1}) is  using  Monte Carlo 3-d lattice simulations via  dimensional reduction \cite{3-d}.  
In dimensional reduction  an effective 3-d theory is constructed perturbatively integrating
out all the massive modes. The characteristics  of the phase transition can
then be determined
 by simulating on the lattice the remaining 3-dimensional
 theory of massless bosonic modes. 
The strength of the lattice simulations is that they can constrain
a whole class of models which are described by the same effective 3-d theory containing a single light scalar 
at the phase transition.

Simulations have also been done for the case where there
is a colored $SU(2)$ singlet among the light/massless
modes at $\phi = 0$ \cite{KetalMSSM}. This
corresponds to the light RH stop scenario that
could make the phase transition strong
enough in the MSSM. The area of MSSM parameter
space where the EWPT is strong enough
was found in the lattice analysis
to be larger than the area found in perturbation theory.
This means that calculating $\phi(T_c)/T_c$ from
the perturbative effective potential is
conservative, so this is
what we will do. We construct the effective potential via dimensional
reduction.  

The effective 3D theory constructed with dimensional reduction
reproduces  the perturbative 4D effective potential results.
The 3-D theory naturally incorporates the effects of resummation and some higher order corrections.
We use the results given in ref. 
\cite{Losada4} for the two-loop finite-temperature
effective potential of the MSSM with a light stop.
 It assumes that
the $b$-quark Yukawa coupling is small
(this restricts the value of  $\tan \beta \lsim 15$), and is calculated in
the limit where all the supersymmetric particles are
heavy ($\sim $ TeV) except for the stops\footnote{ We include scalar boson doublets and singlets 
with masses $m \lsim 2\pi T$, such that the high-temperature expansion is
valid. Neglecting the contributions from all other sfermions and other supersymmetric particles which do not couple strongly to
the Higgs boson is a good approximation. The Higgsino does couple through
the top Yukawa but as it is a fermion it does not strongly  affect the
strength of the phase transition.}. 
 The heavier stop is included in
the high-temperature expansion, rather
than as a TeV-mass particle, because the areas
of CMSSM parameter space we are interested in typically predict
a heavy stop mass $\ga 300$ GeV.

The conclusions of the two-loop perturbative analysis for the MSSM 
\cite{BJLS, CQW, Losada4} are that the sphaleron transitions
are suppressed when the  stop  and  Higgs bosons are light enough. 
However, the maximum possible values of $m_{\tilde{t}_2}$ (the light
stop mass), and $m_h$  for which (\ref{1}) is satisfied depend also
on the trilinear terms and on $\tan \beta$. The presence of 
$h_t \mu H_1^{\dagger} Q_3 U_3^c + h_t A H_2 Q_3 U^c_3$ in
the potential weakens the strength of
the phase transition, because these terms reduce
the contribution of the light stop $\tilde{t}_2$ to $E$ (that is, E decreases
for fixed $m_{\tilde{t}_2}$ as $(A + \mu \cot \beta)$ increases),
it makes the Higgs heavier and increases
the value of $T_c$. These undesirable effects
of trilinears on the survival of the BAU can be partially compensated by decreasing the
singlet stop soft mass $m_{U_3}^2$. So if for fixed
 $(A + \mu \cot \beta)$ one can independently
decrease $m_{U_3}^2$, one
can find a combination of $m_h$, $m_{\tilde{t}_2}$,
 $(A + \mu \cot \beta)$ ( and $\tan \beta$)
such that the phase transition is strong
enough. Of course,
there is a lower bound on  $m_{\tilde{t}_2}$ from direct
searches, which puts an upper bound on the amount of
mixing in the stop sector with which the BAU can be
preserved. However, as we shall see, it is difficult in the CMSSM
to make $m_{U_3}^2$ at the weak scale smaller for fixed 
$(A + \mu \cot \beta)$ , because
it is the $A$ parameter that one is using to run
the soft mass $m_{U_3}^2$ negative (see eqn. \ref{eq:stprge}),
so the  $m_{U_3}^2$ and $(A + \mu \cot \beta)$  are not
independent. In the CMSSM it is therefore 
not trivial to produce a light stop with little left-right mixing.

The physical masses $m_h$ and $m_{\tilde{t}_2}$ for which the
phase transition is sufficiently first order depend mildly
on $\tan \beta$. The Higgs mass
increases with $\tan \beta$, so for larger $\tan \beta$
one needs a lighter stop.  For fixed $ \tan \beta$,
the zero stop mixing case  $(A + \mu \cot \beta)= 0$
provides the least stringent bound on  $m_{\tilde{t}_2}$,
which is of order  $m_{\tilde{t}_2} < m_t$ (as discussed
following eqn \ref{AAA}). The largest possible value of
 the Higgs mass $m_h$
is determined by the largest allowed value of
the mixing parameter; for $m_{Q}=300$ GeV it is $m_{h} \lsim 105$ GeV.
This is because the stop loop contribution to $m_h$
increases with stop mixing.

A two-stage phase
transition can appear 
at finite temperature in certain regions of parameter space. The
fact that the stop is light allows the possibility of tunneling into
a color and charge breaking minimum \cite{BJLS, CQW, Losada4, CMS}, from which  
the Universe would subsequently undergo a transition to the $SU(2)$ broken minimum. The recent analysis
of ref. \cite{CMS} shows that the second phase transition may not take place, thus giving stronger 
constraints on the allowed parameter 
space. This gives a lower bound on the stop mass for every set
of values of the mixing parameter and 
$\tb$.  This lower bound is larger than the direct
experimental search limit on $m_{\tilde{t}_2}$.
In the present paper we do not include this constraint.

We are  interested in an area of CMSSM
parameter space where
the Higgs is light and the singlet stop soft
mass is negative. The one-loop Renormalization Group Equation (RGE)
for this stop mass is \cite{dBoer}
\be
\frac{d m^2_{\tilde{U}_3}}{dt} = \frac{16}{3} \frac{\alpha_3}{4 \pi} M_3^2
 -2 \frac{h_t^2}{16 \pi^2}(m^2_{Q_3} + m_{U_3}^2  + m_{H_2}^2 + A_t^2 )
\label{eq:stprge}
\ee
where $t = ln(m_{GUT}^2/Q^2)$, 
 $M_3$ is the gluino mass, $\alpha_3$ is the strong coupling,
$H_2$ is the Higgs that couples to the top quarks, and 
we have  neglected contributions proportional to $\alpha_1 M_1^2$. 
One can see that the gluino runs the stop mass up as the soft masses
are evolved down from the GUT scale,
but a sufficiently 
large $|A_0|$ at the GUT scale will run 
$ m^2_{\tilde{U}_3} \lsim 0$ near the electroweak scale.
From approximate analytic solutions to the one-loop
RGEs \cite{Steve}, one finds that  $m^2_{\tilde{U}_3} \sim 0$
for $|A_0| \sim 11 |m_{1/2}|$.

In addition to the restrictions imposed on the CMSSM by the
preservation of the baryon asymmetry, we consider both experimental
constraints from LEP particle searches \cite{LEP} and cosmological
constraints on the relic density of the lightest supersymmetric
particle.  Over much of the CMSSM parameter space, the LSP is a bino
$\widetilde B$.  For a gaugino-type neutralino, annihilation in the
early universe occurs primarily via sfermion exchange into fermion
pairs, and in particular into lepton pairs in the CMSSM.  Since the sleptons
become more massive with $m_0$, the neutralino annihilation rate
decreases, and the neutralino relic abundance increases as $m_0$ is
raised, leading to an effective upper bound on the scalar mass
parameter.  An exception occurs when the neutralino mass is close to
half the Higgs or $Z$ mass, so that s-channel annihilation on the
Higgs or $Z$ pole can dominate, and the relic abundance satisfies
$\ohsq\le0.3$ independent of $m_0$.

\begin{figure}[htb]
\hspace*{1.50in}
\epsfig{file=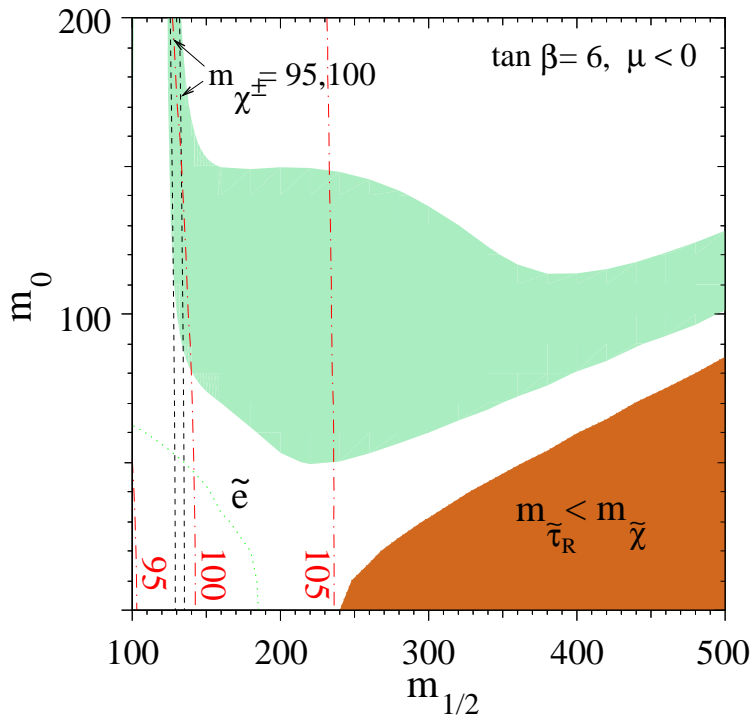,height=3.7in} 
\caption{\label{fig:cmssm}
{\it The light-shaded area is the cosmologically preferred 
region with \protect\mbox{$0.1\leq\ohsq\leq 0.3$}.   
In the dark shaded region, the LSP is
the ${\tilde \tau}_R$, leading to an unacceptable abundance
of charged dark matter.  Also shown are the isomass
contours $m_{\chi^\pm} = 95,100$~GeV (dashed) and $m_h = 95,100$~GeV (dot-dashed),
as well as an indication of the slepton bound (dotted line) from
LEP~\protect\cite{LEP}. }}
\end{figure}

In Fig.~\ref{fig:cmssm}, the cosmological and experimental bounds are
displayed in the $(m_0,\m12)$ plane for the representative choice
$\tb=6$, $A_0=0$, and $\mu<0$.  The light shading denotes the region where the
neutralino relic density lies in the cosmologically preferred range
$0.1\le\ohsq\le0.3$ and is fairly insensitive to $\tb$ for the small
to moderate values of $\tb$ which are relevant.  The upper bound (which provides the boundary at
large $m_0$ and $\m12$) arises from the requirement that the universe
be older than $t_U>12$ Gyr.  The lower limit requires that the relic
neutralinos contribute appreciably to the dark matter of the universe
and is more of a preference that a bound, in contrast to the upper
limit.  The narrow chimney which extends to large $m_0$ at $\m12\sim
120\gev$ is due to s-channel annihilation on the Higgs and $Z$ poles.
In the dark shaded region of Fig.~\ref{fig:cmssm}, the lightest
supersymmetric particle is the stau $\tilde\tau_R$, which is forbidden
by the stringent limits on charged dark matter \cite{ehnos}.  The
shaded region bends away from the line $m_{\tilde\tau_R}=\mchi$ due to
stau annihilation and stau-neutralino coannihilation, which
significantly deplete the neutralino relic abundance for small
neutralino-stau mass splittings \cite{efo}.  We'll see below
that preserving the baryon asymmetry requires $\m12\lsim300\gev$, and
so from Fig.~\ref{fig:cmssm}, $\ohsq\le0.3$ necessitates
$m_0\lsim170\gev$, outside of the pole region.  

The experimental bound on the chargino mass in the CMSSM saturates the
kinematic limit of $\sim95$ GeV, except in a tiny region where the chargino
is just slightly heavier than the sneutrino, which we neglect.  We
display as dashed lines in Fig.~\ref{fig:cmssm} contours of constant
$m_{\tilde\chi^\pm}=95$ and 100 GeV, the latter of which approximates
the chargino mass reach of LEP200.  The chargino bound cuts off most
of the cosmological s-channel pole region.  Light dotted contours
indicate slepton bounds from searches for acoplanar lepton pairs at
LEP183 and do not make inroads into the light shaded region.
Constraints from Higgs searches are most severe at low $\tb$, where
the tree-level Higgs mass $m_h\approx m_Z|\cos{2\beta}|$ is smallest,
and where the experimental lower bounds on the Higgs mass are strongest.  The radiative
corrections to the Higgs mass are large\cite{MSSMHiggs}
and increase logarithmically with the stop masses; thus the Higgs mass
bound may be satisfied for sufficiently large $\m12$, although such
large values may be excluded from other, e.g. cosmological,
considerations \cite{tf1}.  Higgs mass contours are shown as
dot-dashed lines in Fig.~\ref{fig:cmssm}. As $\tb$ is increased, the
tree-level Higgs mass grows, and additionally, the experimental lower
bound on the Higgs mass falls; however, one becomes more susceptible
to the upper bound on the Higgs mass from baryon asymmetry
preservation, as above.

Supersymmetric particles could also
contribute via loop effects to
precision observables. Relevant 
constraints come, for instance, from
the $\rho$ parameter \cite{Djetal}, which receives
contributions when there is
a mass difference  between
two members of an $SU(2)$ doublet. 
In the stop sector, the large top
mass means that the doublet and singlet
stops are in general mixed, i.e. the mass eigenstates
are linear combinations of  the
doublet and singlet states, and so both mass eigenstates contribute
to $\rho$. We need
a light stop $m_{\tilde{t}_2} \sim m_t$
for electroweak baryogenesis; for
this to be consistent with the determination
of $\rho$,
the light stop needs to be mostly
singlet, and the doublet stop heavy enough.
This  happens for some parameters in the CMSSM
because the  RGEs run the soft singlet mass
smaller than the doublet mass,  and 
the doublet-singlet mixing will be small
if ($A + \mu \cot \beta)$ is small.
 It is worth
clarifying that the constraint from the $\rho$ parameter
depends on the amount of mixing.
As the stop mixing angle increases from zero, the lower
bound on the soft mass of the
heavy squark ($\approx $ doublet) {\it decreases}, but
for large enough mixing it increases again \cite{MCrho}.

Another constraint on SUSY models
is the decay $b \rightarrow s \gamma$,
to which a light charged Higgs and/or a light
chargino and squark  can make
substantial contributions \cite{bsg}.
For the area of parameter space we are interested in, the charged
Higgs is moderately heavy (400 --700 GeV) but
we certainly  have a light stop.
The contribution to the $b \rightarrow s \gamma$ amplitude
from charged Higgs-top exchange in SUSY is always negative, as is
the SM amplitude, and 
decreases in magnitude as 
$m_{H^+}$ increases. The chargino-squark contributions are more 
complicated. 
 If the up and charm squark
masses are comparable to that of the doublet stop, as
is the case for our parameters, then the charm squark-gaugino
amplitude plus the up squark-gaugino 
amplitude  will be of opposite
sign to the stop-gaugino contribution (because of CKM unitarity)
 so the up and charm squarks  can
partially cancel  the (heavy doublet) stop  amplitude.
 The squarks also have
Yukawa-strength interactions with the Higgsino component of the chargino,
which are strong for the stops.  This ``stop-Higgsino'' 
 amplitude is proportional to
the singlet-doublet stop mixing, which is small in
the area of parameter space we are interested in. However, 
for our parameters, the sign of the
stop-Higgsino amplitude is $- sgn(\mu)$.  Since
we take $\mu$ positive  so as to cancel against $A$ 
(which is negative in the regions of interest) in the stop
mixing mass, we find that this remaining stop-chargino contribution
is {\it negative} like the SM and charged Higgs amplitudes.
This absence of cancellation between the SUSY
contributions to $b \rightarrow s \gamma$ will make the branching
ratios in our ``allowed for DM and the BAU''  region larger 
than in the SM, and potentially problematic.

\section{Results}

We now present our analysis of the regions of the CMSSM parameter
space which satisfy the experimental and cosmological constraints
outlined above.  In all our numerics, we run the soft mass parameters
and gauge and Yukawa couplings to the electroweak scale with two loop
RGEs for the couplings and gaugino masses and one-loop RGEs for other
masses.  In practice, we find that the stop mass constraint is the
most difficult to satisfy.   This is because the gluino contribution
to the running of the stop mass$^2$ parameters in (\ref{eq:stprge})
tends to drive the light stop mass to values larger than that
permitted by the preservation of the baryon asymmetry.   The terms in 
(\ref{eq:stprge}) proportional to the top Yukawa coupling drive the
stop masses down, and at fixed $\m12$ and $m_0$, we have the option of
varying $A_0$ to help produce a light stop.  We will see that this
requires a fairly large and particular range of values for $A_0$.  Of
course, there is a similar term proportional to the tau Yukawa coupling 
in the RGE for the stau mass$^2$ parameters, and for some ranges of
parameters at medium to large $\tb$, the light stau mass$^2$ can be
driven negative.

\begin{figure}[htb]
\hspace*{-0.5cm}
  \begin{minipage}{6.0cm}
    \epsfig{file=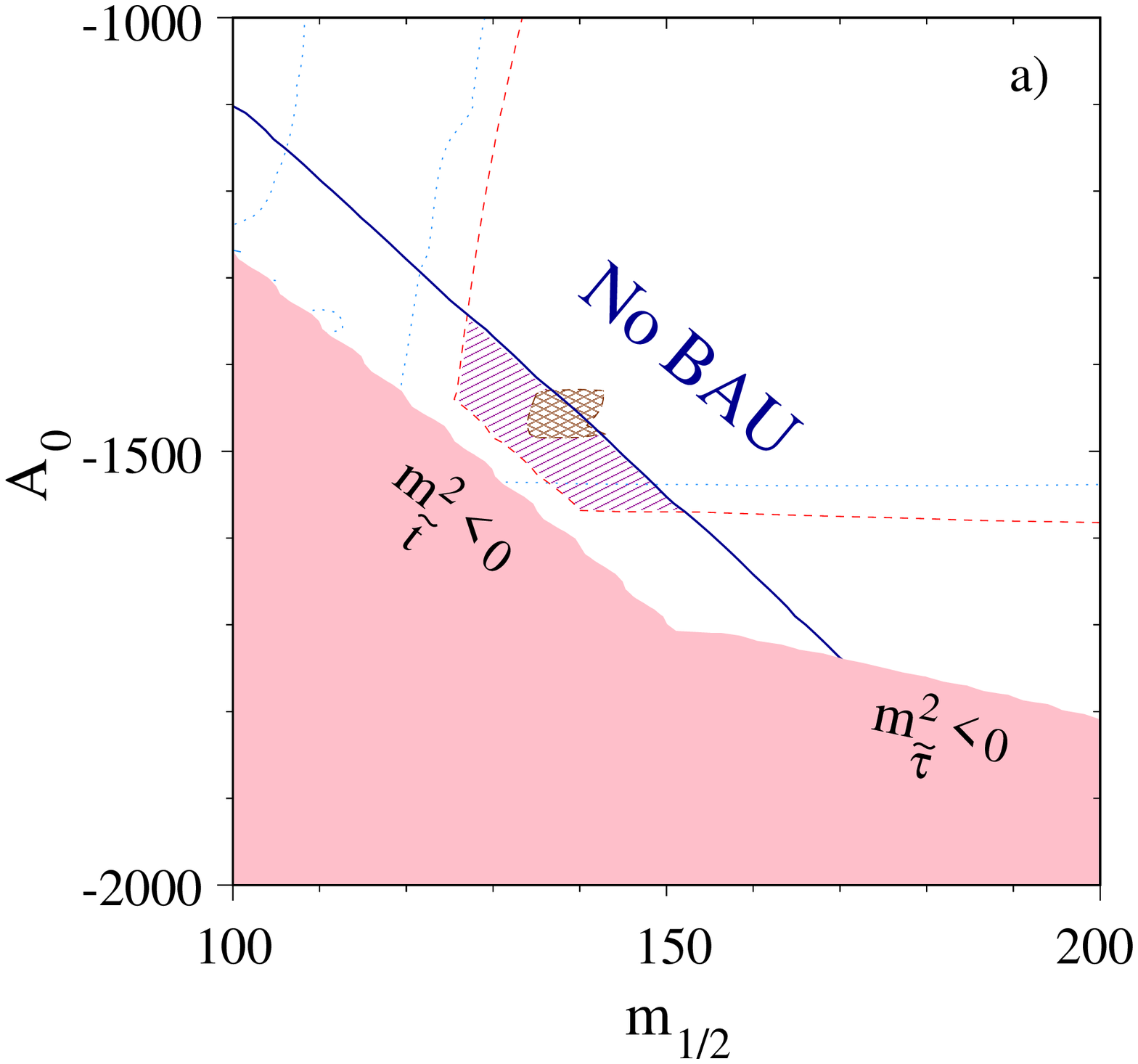,height=3.7in} 
  \end{minipage}
\hspace*{1.6in}
\vspace*{0.1cm}
  \begin{minipage}{6.0cm}
    \epsfig{file=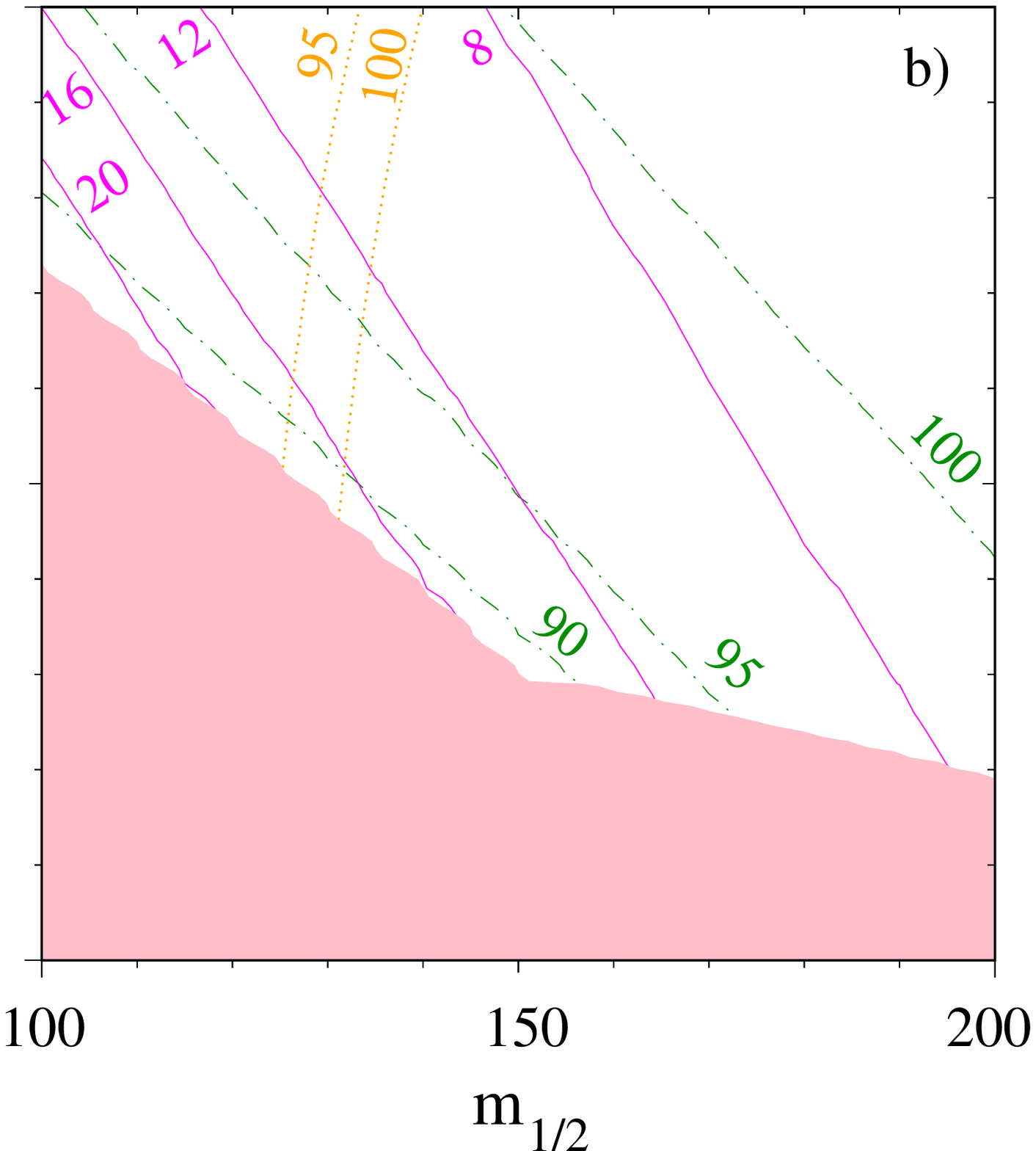,height=3.66in} 
  \end{minipage}
\caption{\label{fig:ma12.45n}
{\it In a), the region of mSUGRA parameter space satisfying the BAU,
  experimental and cosmological constraints described in the text, for $\tb=12,
  m_0=145$ GeV and $\mu>0$, is marked by diagonal shading.  Above the solid line, sphaleron processes 
  wipe out the baryon asymmetry.  The light shaded region contains
  tachyons.  The dark hatched  region has $\ohsq>0.3$.  The dotted lines 
  are contours of constant $\ohsq=0.1$.  Contours of constant Higgs
mass (dot-dashed), chargino mass (dotted) and
$\delta\rho\times10^{4}$ (solid) are displayed in b).}}
\end{figure}

We now fix $\tb=12$ and $m_0=145\gev$ and plot the combined set of
experimental and theoretical constraints in the $(\m12,A_0)$ plane. 
As we're  interested in the region where the doublet-singlet
stop mixing angle is reasonably small, we concentrate here on
the area of parameter space where the sign of
$\mu$ is opposite to that of $A$, to allow
cancellation  in the off-diagonal elements of
(\ref{AAA}).   Also, the presence of a quasi-fixed point for $A_t$ at low $\tb$ 
tends to drive $A_t$ to positive values in the vicinity of $2\m12$\cite{irqfp}, which tends
to be much larger than $|\mu|\cot\beta$.   Smaller $|A_t|$ can be
generated at moderate $\tb$ by taking $A_0$ large and negative.
Accordingly, in Fig.~\ref{fig:ma12.45n}a, we consider negative $A_0$ and positive
$\mu$.  The light shaded region is excluded because it contains either a
tachyonic stop or stau.
The dashed contour combines the experimental bounds from chargino, stop 
and Higgs searches with the condition that the LSP be the neutralino,
to avoid an excess of charged dark matter \cite{ehnos}.  In
particular, for the Higgs 
bound, we take  91 GeV at $\tb=4$ and 83 GeV at
$\tb=6-12$ \cite{delphiconf}. In Fig.~\ref{fig:ma12.45n}a, the vertical left side of the dashed contour is due
to the chargino bound, the diagonal piece which parallels the light
shaded region is due to the stop bound, and the horizontal piece
is the line $\mchi=m_{\tilde\tau_R}$ \footnote{We have not included
  the new preliminary bounds on the stau mass\cite{LEP}, which may
  slightly reduce the BAU allowed region.}.  The experimental Higgs
constraint does not provide a useful bound for this value of $\tb$.
The solid line gives the BAU constraint:  above the solid line,
sphaleron processes wash out the baryon asymmetry, while below the
solid line the baryon asymmetry is preserved.  
In the figures we present, the BAU boundary is typically set by the condition that
the stop mass be sufficiently light.
Lastly, the dark hatching marks the
region where the neutralino relic density violates the cosmological upper bound
$\ohsq\le0.3$.  The region which is
allowed by all of the experimental and cosmological constraints is then
highlighted by diagonal shading.
We show as the dotted line the lower boundary of the
cosmologically preferred region, $\ohsq=0.1$.   Note that it bends
away from the $\mchi=m_{\tilde\tau_R}$ contour, as stau-neutralino
coannihilation drives the relic abundance to small values for close
stau-neutralino mass degeneracy.  Notice also the chimney at
$\m12\sim110-130\gev$, which is an analogous s-channel pole structure
to that in Fig.~\ref{fig:cmssm}.

\begin{figure}[htb]
\hspace*{-0.5cm}
  \begin{minipage}{6.0cm}
    \epsfig{file=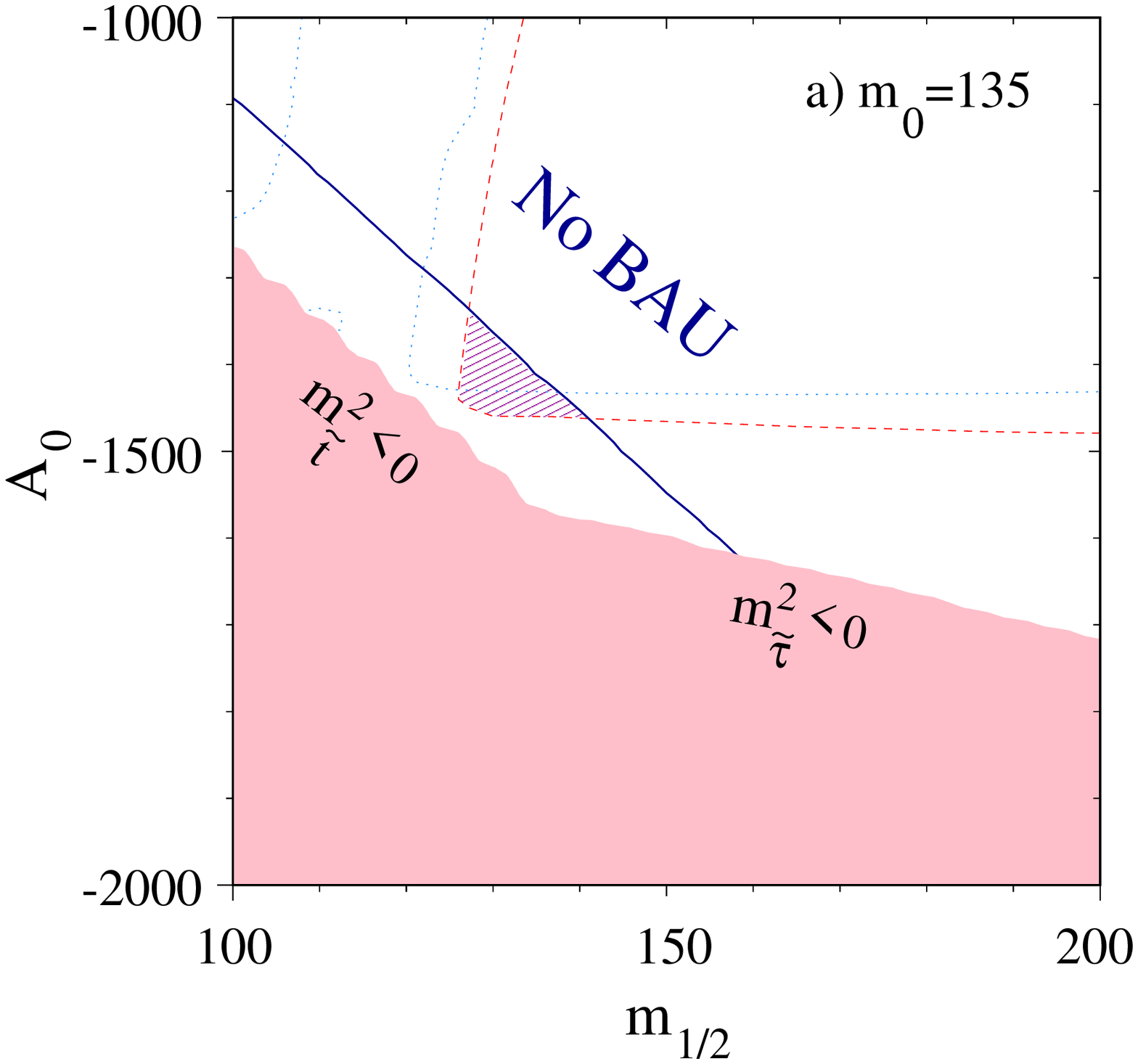,height=3.7in} 
  \end{minipage}
\hspace*{1.6in}
\vspace*{0.1cm}
  \begin{minipage}{6.0cm}
    \epsfig{file=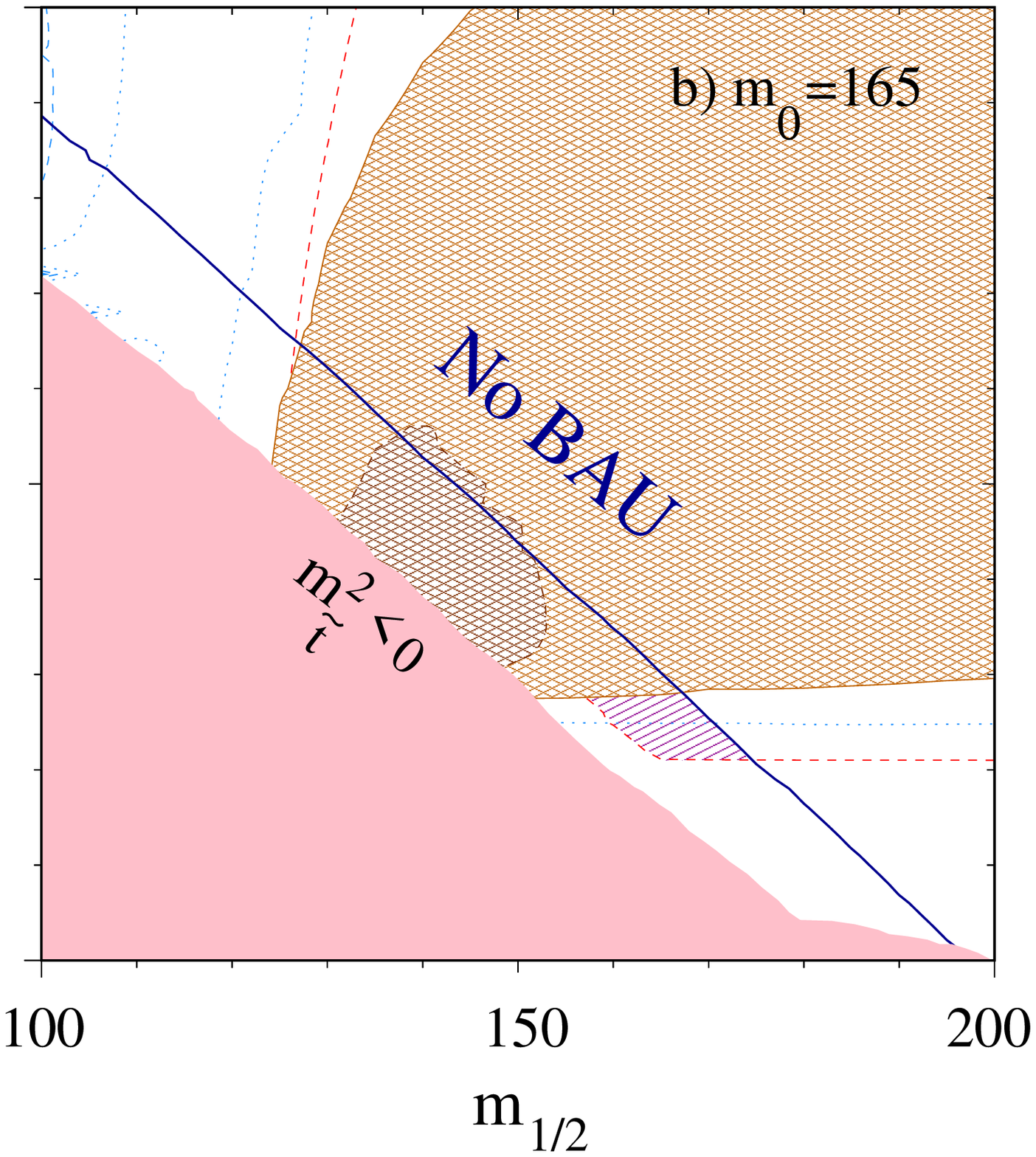,height=3.68in} 
  \end{minipage}
\caption{\label{fig:xm0}
 {\it The same as Fig.~\ref{fig:ma12.45n}a, but for a)~$m_0=135\gev$ and 
  b)~$m_0=165\gev$}.  In b) the dark hatched region has $\ohsq>0.4$
and the light hatched region has $0.3<\ohsq<0.4$.}
\end{figure}

\begin{figure}[htb]
\hspace*{-0.5cm}
  \begin{minipage}{6.0cm}
    \epsfig{file=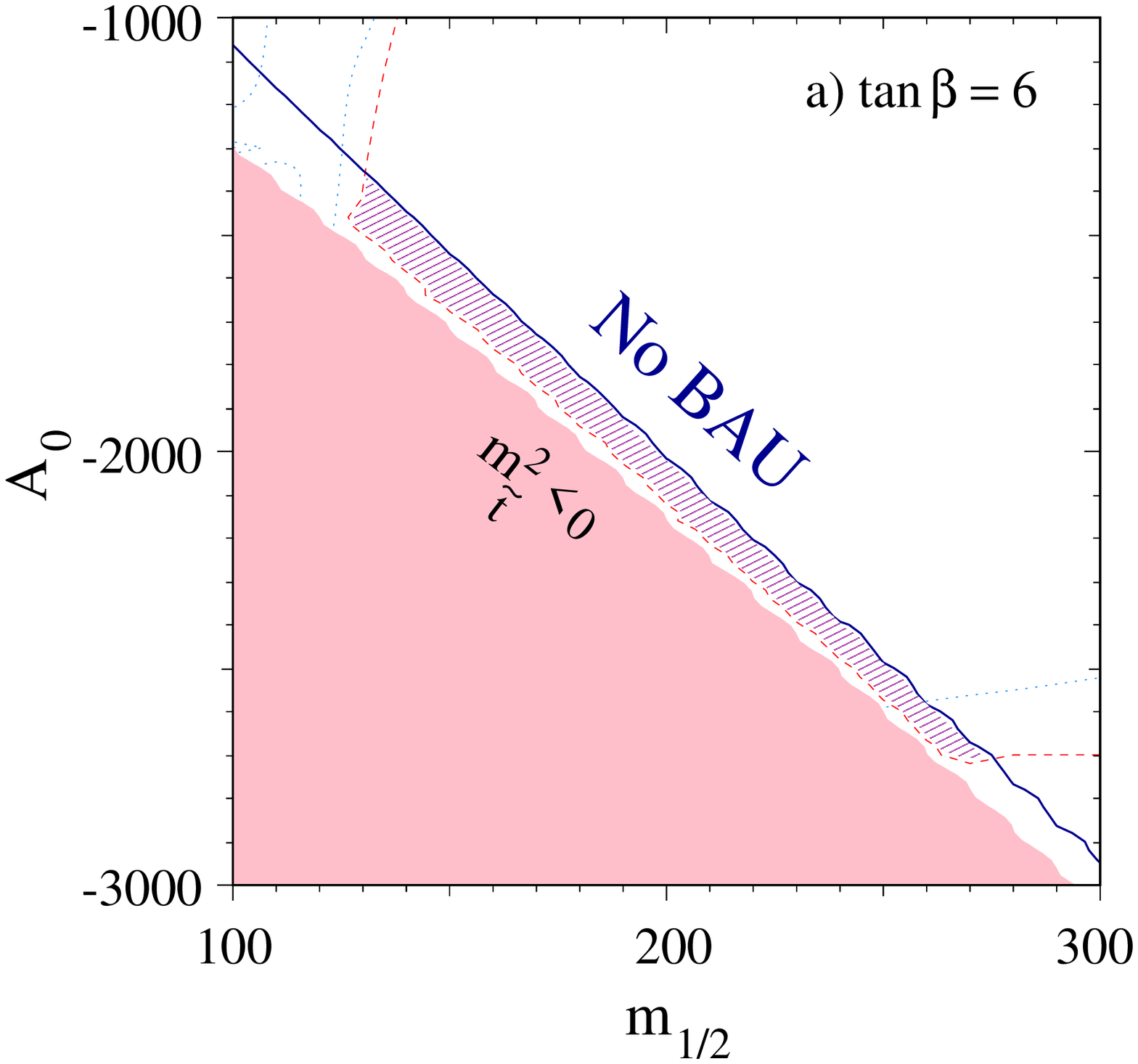,height=3.7in} 
  \end{minipage}
\hspace*{1.6in}
\vspace*{0.1cm}
  \begin{minipage}{6.0cm}
    \epsfig{file=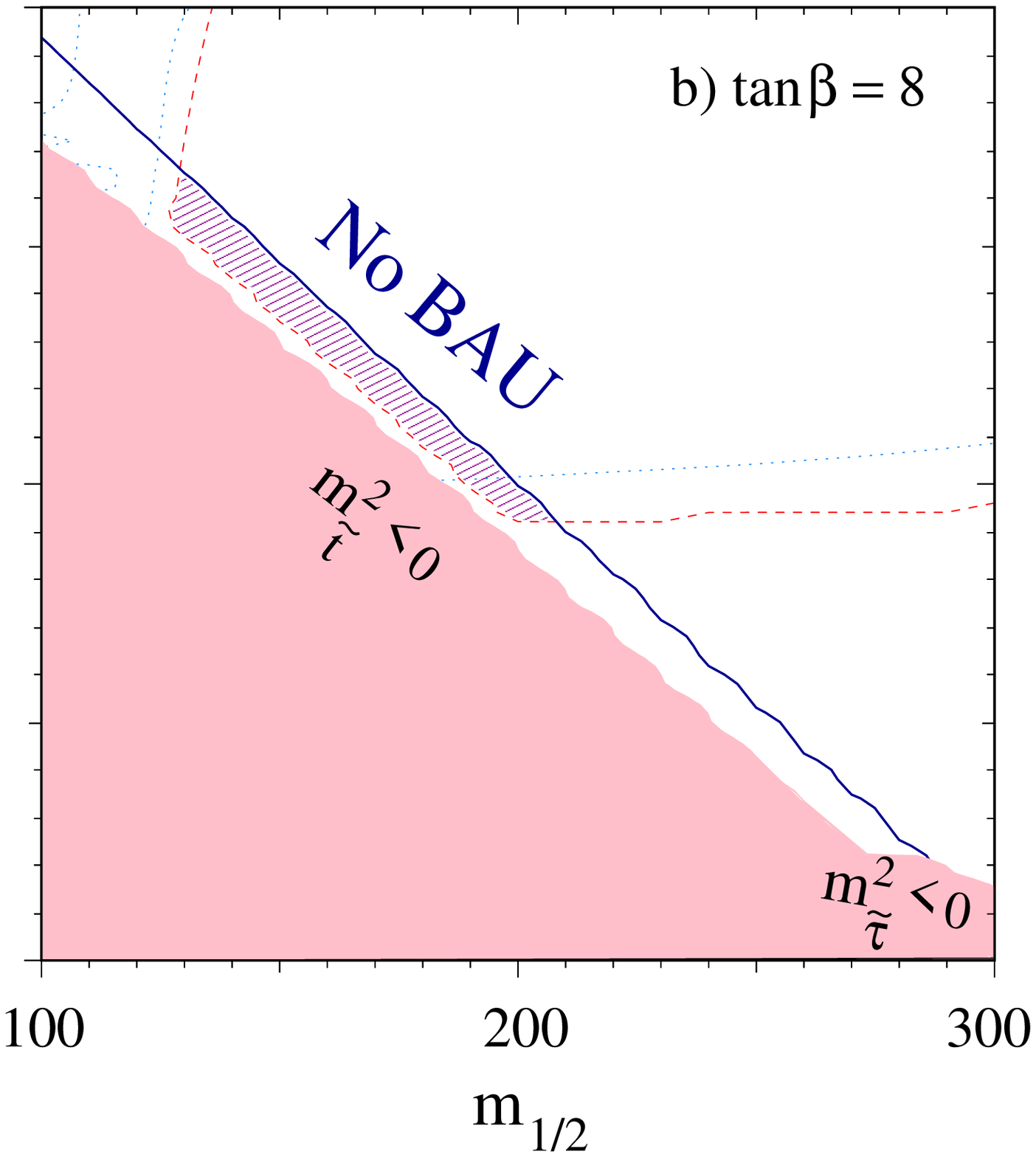,height=3.68in} 
  \end{minipage}
\caption{\label{fig:xtb}
{\it The same as Fig.~\ref{fig:ma12.45n}a, but for a) $\tb=6$ and
  b) $\tb=8$.}  $\ohsq$ is less than 0.3 everywhere in the plots.}
\end{figure}

In Figs.~\ref{fig:xm0} and \ref{fig:xtb}, we show the variation of BAU
allowed region with $m_0$ and $\tb$.  In Fig.~\ref{fig:xm0}a and
\ref{fig:xm0}b, we've taken $m_0=135\gev$ and $165\gev$, respectively,
for $\tb=12$.  For $m_0=135\gev$, the staus are lighter vis-a-vis the
neutralinos, the horizontal piece of the dashed contour moves upward,
and the combined chargino and LSP constraint almost wipe
out the BAU allowed region entirely.  Recall from Fig.~\ref{fig:cmssm}
and the discussion in Section 2 that larger $m_0$ implies heavier
sleptons, and a higher relic neutralino abundance, while lower $m_0$
implies a lower neutralino relic density.  This is evident in
Fig.~\ref{fig:xm0}a, where no contours of $\ohsq=0.3$ are visible.  In
Fig.~\ref{fig:xm0}b, the staus are heavier vis-a-vis the neutralinos,
and the combined experimental and BAU constraints permit a larger
parameter region.  However, here the relic densities are larger, and
the constraint $\ohsq<0.3$ excludes the bulk of the BAU allowed
region, leaving a small region near the line $\mchi=m_{\tilde\tau_R}$,
as above.  

In Fig.~\ref{fig:xtb}, we show the BAU allowed regions for $\tb=6$ and
$8$, at $m_0=125\gev$.  Because the stau mass runs less for smaller
$\tb$, the horizontal piece of the dashed exclusion contour is moved
to lower values of $A_0$, opening up more BAU allowed parameter space.
One can also take a smaller $m_0$, so that the relic density constraint
is easily satisfied.  As in the case $\tb=12$, the allowed range of $m_0$ is
limited on the low side by the requirement that the LSP be the
neutralino, and on the high side by the relic density constraints.
For $\tb\le4$, the tighter experimental Higgs mass bound excludes the entire
BAU allowed region.  Overall, we find that the range
$5\lsim\tb\lsim12$ yields the largest BAU regions which satisfy the
experimental and cosmological constraints.   For simplicity, we've
applied a single Higgs mass constraint (corresponding to large stop
mixing) for all $\m12$ and $A_0$ at fixed $\tb$.  Of course in the BAU allowed
regions, the stop mixing is in fact small, and the experiment bounds for $\tb\le6$
are consequently tighter than those we have imposed and exclude all
but a sliver of the BAU allowed region in  Fig.~\ref{fig:xtb}a.  The
other figures are unaffected.

We emphasize that the
signs of $A_0$ and $\mu$ were selected by the requirement that there
be cancellations in the off-diagonal element of the stop mass matrix,
leading to a small stop mixing angle.  This requires not only that
$A_t$ and $\mu$ have opposite signs, but also that $|A_t|$ be
considerably below its quasi-fixed point value, which can only be
achieved for $A_0$ large and negative.   It is also possible to get small
stop mixing for $\mu<0$, since one can get small, positive $A_t$ for large negative $A_0$.
In fact, BAU regions do exist for $\mu<0$, but they are considerably smaller even than the 
somewhat narrow allowed regions of Figs.~\ref{fig:ma12.45n}-\ref{fig:xtb}.

We compute the one-loop 
stop contribution to $\Delta \rho$ from \cite{Djetal}.
Contributions from other
SUSY particles are not included
on our plots; from \cite{Chanetal}
we expect the sleptons and ``-inos''
to give at most $\Delta \rho < 0.0009$. 
Following \cite{Djetal,Alt}, we
would like the stop contribution
to be less than the one-sigma
error on $\epsilon_1$, which
is \cite{PDB} 0.0012. All the 
supersymmetric contributions
to $\Delta \rho$ should therefore
fit comfortably within the two
sigma error bars. In
Fig.~\ref{fig:ma12.45n}b, we show contours of constant $\Delta \rho$,
along with chargino and Higgs mass contours, to demonstrate how the BAU
 allowed region will be further constrained by improved
experimental bounds in the future.   For this value of $\tb$, the
entire BAU region is already excluded by the requirement
$\Delta\rho<0.0012$.  For $\tb=6$ and 8, by contrast, the entire BAU
allowed regions have  $\Delta\rho<0.0010$ and 0.0012, respectively.

To estimate the $b \rightarrow s \gamma$ branching ratio
for the parameter space we are interested in,
 we have calculated the
LO contributions to $C_7(m_W)$ and $C_8(m_W)$ following \cite{copy},
and the  branching ratio from
formulae in \cite{KN}, who include NLO corrections in the running
from $m_W$ to $m_b$. The branching ratios we find
are   large :$ BR \sim 4.1$---$6.7 \times 10^{-4}$.
CLEO measures $3.15 \pm 0.35 \pm 0.32 \pm 0.26 \times 10^{-4}$ \cite{CLEO}.
If we impose the requirement that the branching ratio
we calculate be smaller than the CLEO $95\%$ C.L. measurement
($B.R. < 4.5 \times 10^{-4}$ \cite{CLEO}), 
then we find that the $\tan \beta = 6$ and 
8 ``allowed for baryogenesis and dark matter''  regions 
satisfy this requirement
($BR\sim 4 \times 10^{-4}$). However, over
most of the $\tan \beta = 12$ region,
we find branching ratios of order 6 -- 7 $\times 10^{-4}$,
which is too large.
The branching ratio is very sensitive to the charm and up squark masses,
 because these
 squarks contribute to $C_7(m_W$) with the opposite sign from the stop, and
so even a relatively small relaxation of the the CMSSM boundary conditions
for the squarks can have a significant impact on  the rate for $b 
\rightarrow s \gamma$.
 For instance,
decreasing the charm squark mass by 100 GeV ($15-25 \%$)
decreases the branching ratio we calculate below the CLEO 95$\%$
bound.  A NLO \cite{Paolo}  calculation would be required 
to reliably calculate $b \rightarrow s \gamma$
in our large $\tan \beta$ region, because the
NLO corrections can be  significant when there
are cancellations between different contributions.
In summary,  although  $BR( B \rightarrow X_s \gamma$) is
greater than in the Standard Model,
it is not prohibitively large, at least for $\tan \beta = 6, 8$,
because 1) the charged Higgs is moderately heavy, 2)
the charm squark and gaugino partially cancel the stop 
and gaugino contribution, and 3) the stop and Higgsino
contribution is proportional to the stop mixing,
which is small.

Lastly, we note that if a Higgs is not observed at LEP200, the increased Higgs mass bounds
will exclude  most of the remaining BAU allowed parameters space in
the CMSSM.    If any BAU region remains, it will be at large
$\tb\ga10$, where the Higgs bound can fall below $\sim 95$GeV.   The
BAU allowed regions in Figs.\ref{fig:xtb} have Higgs masses between
$\sim 88-92$ GeV and will be forbidden if the Higgs is not seen before
the end of LEP running.

\section{Conclusions}
In the CMSSM, we have found the area of parameter
space consistent with present experimental limits which  predicts
a neutralino relic density $ 0.1\la\Omega h^2 \la 0.3$,
and where sphaleron processes do not wipe out the Baryon Asymmetry of
the Universe. 
The allowed regions correspond
to large  $|A_0|$, because this ensures
that the RH stop soft mass is small,
as required by the baryogenesis constraint.
They also occur for negative $A_0$ and positive
$\mu$,  because this
allows cancellations between $\mu$ and $A$
in the off-diagonal elements of
the squark mass matrix (\ref{AAA}), which is  dictated
by baryogenesis, as well.  There also exist allowed regions at
negative $A_0$ and negative $\mu$, but these are quite tiny, even
compared to the small regions we find for positive $\mu$.
We find that the current Higgs mass constraints exclude
all BAU allowed regions for $\tb\la5$, and for $\tb\ga12$, the BAU
allowed regions become very constrained.  
We find the branching ratio for $ b \rightarrow s \gamma$
in the allowed regions to be larger than the SM, but consistent with
present CLEO data for smaller values of $\tan \beta$.
Lastly, we find that if  LEP200 should not find a Higgs, almost all of 
the currently BAU allowed regions will be excluded, with the possible
exception of those at the largest $\tb$, where the Higgs mass bounds
will be weakest.

\paragraph{Acknowledgments}
S.D. would like to thank Paolo Gambino for
clear explanations, and Steve Abel and Gian Giudice  for
useful conversations.  The work of T.F. was supported in part by DOE   
grant DE--FG02--95ER--40896, and in part by the University of Wisconsin  
Research Committee with funds granted by the Wisconsin Alumni Research  
Foundation.

\end{document}